\documentclass[twocolumn,aps,prd,showpacs]{revtex4}
\usepackage{amssymb}

\newcommand{\Theorem}[2]{\medskip\noi {\bf #1. \ }{\sl #2}\medskip}

\newcommand{\sect}[1]{Sec.\,#1}
\def\nq{\hspace*{-1em}}

\def\nhh{\hspace*{-0.3em}}
\def\cm{\hspace*{1cm}}
\def\inch{\hspace*{1in}}

\def\noi{\noindent}

\def\Jl#1#2{{\it #1\/} {\bf #2},\ }

\def\ApJ#1 {\Jl{Astroph. J.}{#1}}
\def\CQG#1 {\Jl{Class. Quantum Grav.}{#1}}
\def\DAN#1 {\Jl{Dokl. AN SSSR}{#1}}
\def\GC#1 {\Jl{Grav. \& Cosmol.}{#1}}
\def\GRG#1 {\Jl{Gen. Rel. Grav.}{#1}}
\def\JETF#1 {\Jl{Zh. Eksp. Teor. Fiz.}{#1}}
\def\JETP#1 {\Jl{Sov. Phys. JETP}{#1}}
\def\JHEP#1 {\Jl{JHEP}{#1}}
\def\JMP#1 {\Jl{J. Math. Phys.}{#1}}
\def\NPB#1 {\Jl{Nucl. Phys.}{B\ #1}}
\def\NP#1 {\Jl{Nucl. Phys.}{#1}}
\def\PLA#1 {\Jl{Phys. Lett.}{#1A}}
\def\PLB#1 {\Jl{Phys. Lett.}{#1B}}
\def\PRD#1 {\Jl{Phys. Rev.}{D\ #1}}
\def\PRL#1 {\Jl{Phys. Rev. Lett.}{#1}}

\def\al{&\nhh}
\def\lal{&&\nq {}}
\def\beq{\begin{equation}}
\def\eeq{\end{equation}}
\def\bear{\begin{eqnarray}}
\def\bearr{\begin{eqnarray} \lal}
\def\ear{\end{eqnarray}}
\def\eq{Eq.\,}
\def\eqs{Eqs.\,}

\def\nn{\nonumber\\ {}}

\def\nnn{\nonumber\\ \lal }

\def\yy{\\[5pt] {}}

\def\eql{\al =\al}

\def\dst{\displaystyle}
\def\tst{\textstyle}
\def\fracd#1#2{{\dst\frac{#1}{#2}}}
\def\fract#1#2{{\tst\frac{#1}{#2}}}
\def\Half{{\fracd{1}{2}}}
\def\half{{\fract{1}{2}}}

\def\e{{\rm e}}
\def\d{\partial}

\def\diag{\mathop{\rm diag}\nolimits}

\def\const{{\rm const}}
\def\eps{\varepsilon}

\def\GR{general relativity}
\def\sph{spherically symmetric}
\def\ssph{static, spherically symmetric}

\def\bh{black hole}
\def\bhs{black holes}
\def\wh{wormhole}
\def\whs{wormholes}
\def\bw{brane world}
\def\bwd{brane-world}
\def\asflat{asymptotically flat}

\def\Sch{Schwarzschild}
\def\dS{de Sitter}
\def\AdS{anti-de Sitter}
\def\RN{Reissner--Nordstr\"om}

\def\thd{\fract 13}

\def\mn{_{\mu\nu}}

\def\mN{_\mu^\nu}

\def\eff{^{\rm \,eff}}
\def\prad{p_{\rm rad}}
\def\schd{\biggl(1-\frac{2m}{r}\biggr)}

\def\cA{{\cal A}}
\def\cK{{\cal K}}
\def\cR{{\cal R}}

\def\N{{\mathbb N}}
\def\R{{\mathbb R}}
\def\S{{\mathbb S}}
\def\Z{{\mathbb Z}}

\begin{document}

\title {On a general class of brane-world black holes}
\author{K.A. Bronnikov}
\affiliation
  {VNIIMS, 3-1 M. Ulyanovoy St., Moscow 117313, Russia;\\
   Institute of Gravitation and Cosmology, PFUR,
        6 Miklukho-Maklaya St., Moscow 117198, Russia}
   \email {kb@rgs.mccme.ru}

\author{Heinz Dehnen}
\affiliation
  {Fachbereich Physik, Universit\"at Konstanz, Postfach M 677 78457
   Konstanz, Germany}
   \email {Heinz.Dehnen@uni-konstanz.de}

\author{V.N. Melnikov}
\affiliation
  {VNIIMS, 3-1 M. Ulyanovoy St., Moscow 117313, Russia;\\
   Institute of Gravitation and Cosmology, PFUR,
        6 Miklukho-Maklaya St., Moscow 117198, Russia}
   \email {rgs@com2com.ru}

\begin{abstract}
    We use the general solution to the trace of the 4-dimensional
    Einstein equations for \ssph\ configurations as a basis for finding
    a general class of \bh\ (BH) metrics, containing one arbitrary function
    $g_{tt} = A(r)$ which vanishes at some $r = r_h > 0$, the horizon radius.
    Under certain reasonable restrictions, BH metrics are found with or
    without matter and, depending on the boundary conditions, can be
    \asflat\ or have any other prescribed asymptotic. It is shown that
    our procedure generically leads to families of globally regular BHs with
    a Kerr-like global structure as well as symmetric wormholes.
    Horizons in space-times with zero scalar curvature are shown to be either
    simple or double. The same is generically true for horizons inside a
    matter distribution, but in special cases there can be horizons of any
    order. A few simple examples are discussed. A natural application of the
    above results is the \bw\ concept, in which the trace of the 4D gravity
    equations is the only unambiguous equation for the 4D metric, and its
    solutions can be continued into the 5D bulk according to the embedding
    theorems.
\end{abstract}
\maketitle

\section{Introduction \label{s1}}

    The \bw\ concept, which describes our four-dimensional world as a
    surface (brane), supporting all or almost all matter fields and embedded
    in a higher-dimensional space-time (bulk), leads to a great variety of
    models both in the cosmological context and in the description of local
    self-gravitating objects (see, e.g., \cite{bra-rev}
    for reviews and further references). In particular, \bh\ (BH) physics
    on the brane turns out to be considerably richer than in \GR, though
    only a few special examples of \bwd\ BHs have been considered in
    detail by now \cite{bra-bh1}--\cite{casad-qu} (see also references
    therein).  Thus, in the \sph\ vacuum case, in addition to \Sch\ BHs
    (which lead to a black-string singularity in the bulk
    \cite{bra-bh1,kanti02}), there are BHs non-singular on the brane
    \cite{casad01} and having a pancake-shaped event horizon in the bulk
    \cite{casad02}; some of them have been shown to possess unusual quantum
    properties potentially observable on the brane \cite{casad-qu}.

    Most of the results have been obtained in the simplest framework:
    a single brane in a $\Z_2$-symmetric 5-dimensional, asymptotically \AdS\
    bulk, with all fields except gravity confined on the brane. It is the
    so-called RS2 framework, generalizing the second model suggested by
    Randall and Sundrum, with a single Minkowski brane in an \AdS\ bulk
    \cite{ransum2}. Let us also adhere to this class of models.

    The gravitational field on the brane is then described by
    the modified Einstein equations derived by Shiromizu, Maeda and Sasaki
    \cite{maeda99} from 5-dimensional gravity with the aid of the Gauss and
    Codazzi equations \footnote
        {The sign conventions are as follows: the metric signature
         $(+{}-{}-{}-)$; the curvature tensor $R^{\sigma}{}_{\mu\rho\nu} =
    	 \d_\nu\Gamma^{\sigma}_{\mu\rho}-\ldots$, so that, e.g., the Ricci
    	 scalar $R > 0$ for de Sitter space-time, and the stress-energy
    	 tensor (SET) such that $T^t_t$ is the energy density.}:
\bear
    G\mN = - \Lambda_4\delta\mN -\kappa_4^2 T\mN
	        - \kappa_5^4 \Pi\mN - E\mN,                     \label{EE4}
\ear
    where $G\mN = R\mN - \half \delta\mN R$ is the 4D Einstein tensor,
    $\Lambda_4$ is the 4D cosmological constant expressed in terms of
    the 5D cosmological constant $\Lambda_5$ and the brane tension $\lambda$:
\beq
    \Lambda_4 = \Half \kappa_5^2
    \biggl(\Lambda_5 + \frac{1}{6} \kappa_5^2\lambda^2\biggr);  \label{La4}
\eeq
    $\kappa_4^2 = 8\pi G_N = \kappa_5^4 \lambda/(6\pi)$ is the 4D
    gravitational constant ($G_N$ is Newton's constant of gravity); $T\mN$
    is the SET of matter confined on the brane; $\Pi\mN$ is a tensor
    quadratic in $T\mN$, obtained from matching the 5D metric across the
    brane:
\beq                                                             \label{Pi_}
    \Pi\mN = \fract{1}{4} T_\mu^\alpha T_\alpha^\nu - \half T T\mN
           - \fract{1}{8} \delta\mN
    \left( T_{\alpha\beta} T^{\alpha\beta} -\thd T^2\right)
\eeq
    where $T = T^\alpha_\alpha$; lastly, $E\mN$ is the ``electric'' part of
    the 5D Weyl tensor projected onto the brane: in proper 5D coordinates,
    $E\mn = \delta_\mu^A \delta_\nu^C {}^{(5)} C_{ABCD} n^B n^D$ where
    $A, B, \ldots$ are 5D indices and $n^A$ is the unit normal to the brane.
    By construction, $E\mN$ is traceless, $E_\mu^\mu = 0$ \cite{maeda99}.

    Other characteristics of $E\mN$ are unknown without specifying the
    properties of the 5D metric, hence the set of equations (\ref{EE4})
    is not closed. In isotropic cosmology this leads to an additional
    arbitrary constant in the field equations, connected with the density of
    ``dark radiation'' \cite{bra-rev}. For \ssph\ systems to be discussed
    in the present paper, this freedom is expressed in the existence of one
    arbitrary function of the radial coordinate. Despite this arbitrariness,
    the trace of \eqs (\ref{EE4}) may be integrated in a general form
    \cite{bwh1,vis-wilt}.

    Our interest here is in selecting a general class of \ssph\ BH solutions
    to \eqs (\ref{EE4}) without specifying $E\mN$. In particular examples we
    mostly deal with \asflat\ vacuum solutions, such that $\Lambda_4 = T\mN
    =0$, but the BH construction procedure is formulated in the general case
    when both matter and the cosmological constant are present and the
    space-time asymptotic properties are not specified.

    We will not discuss the possible bulk properties of models in question
    and only note that the existence of the corresponding solutions to the
    higher-dimensional equations of gravity (in our case, the 5D vacuum
    Einstein equations with a cosmological term) is guaranteed by the
    Campbell-Magaard type embedding theorems \cite{campmaag}. A recent
    discussion of these theorems, applied, in particular, to \bw\ scenarios,
    and more references can be found in Ref.\,\cite{wess03}.

    The paper is organized as follows. In \sect \ref{s2} we present some
    common relations and the general solution to the trace of \eqs
    (\ref{EE4}), containing an arbitrary generating function $A(r)$.

    In \sect \ref{s3} we analyze the properties of the metric near a Killing
    horizon in a \ssph\ space-time described by the general solution. A
    conclusion of general significance is that a space-time with $R \equiv
    0$ can only have horizons of orders one (simple, like \Sch's) and two
    (double, as in the extremal \RN\ metric) and no higher. This analysis is
    used for formulation of two BH construction algorithms. It is shown that
    a generic choice of $A(r)$ leads to a one-parameter family of solutions
    which, in a certain range of the parameter (integration constant) $C$,
    unifies globally regular non-extremal BHs with a Kerr-like causal
    structure, extremal BHs and symmetric \whs. Singular non-extremal BHs
    can be found outside this range of $C$.

    \sect \ref{s4} contains some simple examples, illustrating different
    features of the present formalism. Examples 1 and 3 reproduce already
    known BH solutions from the viewpoint of our algorithms. Example 2 is a
    BH solution with zero \Sch\ mass, illustrating violation of Thorne's hoop
    conjecture possible in a \bw. Example 4 shows that well-behaved special
    solutions can be found even for such choices of $A(r)$ that the trace
    equation (\ref{master}) has a singular point. Example 5 illustrates the
    smoothness properties of some BH metrics at the horizon in different
    coordinate frames. \sect \ref{s5} is a discussion.

    We will assume that all relevant functions are analytic unless otherwise
    explicitly indicated. The symbol $\sim$, as usual, connects quantities
    of the same order of magnitude in a certain limit.

\section{The general solution \label{s2}}

    The general \ssph\ metric in 4 dimensions in the curvature coordinates
    has the form
\beq                                                          \label{ds-r}
	ds^2 = A(r) dt^2 - \frac{dr^2}{B(r)} - r^2 d\Omega^2
\eeq
    where $d\Omega^2 = d\theta^2 + \sin^2 \theta\,d \phi^2$ is the linear
    element on a unit sphere.

    Let us write down the scalar curvature and the
    Kretschmann scalar for the metric (\ref{ds-r}):
\bear
    R \eql \frac{2}{r^2}(1-B)
\nnn\nq
	-B \biggl[ \frac{A_{rr}}{A} - \frac{A_r^2}{2A^2}
	        + \frac{A_r B_r}{2AB}
        + \frac{2}{r} \biggl(\frac{A_r}{A}+\frac{B_r}{B}\biggr)\biggr];
\yy
        \cK \eql R\mn{}^{\rho\sigma}R_{\rho\sigma}{}^{\mu\nu}
	        = 4K_1^2 + 8K_2^2 + 8K_3^2 + 4K_4^2,
\nn
	K_1 \eql \frac{B}{4}
	\biggl(\frac{2AA_{rr}-A_r^2}{A^2}+\frac{A_r B_r}{AB}\biggr),
\nn
	K_2 \eql \frac{B}{2r}\,\frac{A_r}{A},
\qquad
	K_3 = \frac{B_r}{2r},
\qquad
	K_4 = \frac{1 - B}{r^2},                           \label{Kr}
\ear
    where the subscript $r$ means $d/dr$. The finiteness of $\cK$ is a
    natural regularity criterion for the geometries to be discussed
    since $\cK$ is a sum of squares of all components $R\mn{}^{\eps\sigma}$
    of the Riemann tensor for the metric (\ref{ds-r}), therefore $\cK <
    \infty$ is a necessary and sufficient condition for the finiteness of
    all algebraic curvature invariants. Meanwhile, $\cK$ is finite if and
    only if all $K_i$ defined in (\ref{Kr}) are finite.

    If we treat \eqs (\ref{EE4}) as the conventional Einstein equations with
    an effective SET ${T\mN}\eff$, i.e., $G\mN = -\kappa_4^2 {T\mN}\eff$,
    then the effective density $\rho\eff$, radial pressure $\prad\eff$ and
    tangential pressure $p_\bot\eff$ are expressed in terms of $A$ and $B$
    as follows:
\bear
     G^t_t \eql -\kappa_4^2 \rho\eff = \frac{B-1}{r^2} + \frac{B_r}{r},
\nn
     G^r_r \eql \kappa_4^2 \prad\eff = \frac{B-1}{r^2} + \frac{BA_r}{A r},
\nn
    G^\theta_\theta = G^\phi_\phi \eql \kappa_4^2 p_\bot\eff    \label{T_eff}
	= \frac{B}{4}\biggl[
	\frac{2A_{rr}}{A} - \frac{A_r^2}{A^2}
\nnn \cm
	- \frac{A_rB_r}{AB}
	+ \frac{2}{r} \biggl(\frac{A_r}{A} + \frac{B_r}{B}\biggr)\biggr].
\ear

    The only combination of the Einstein equations (\ref{EE4}) in a brane
    world written unambiguously without specifying $E\mN$, is their trace:
\beq
	R = 4 \Lambda_4 + \kappa_4^2 T^\alpha_\alpha
	        + \kappa_5^4 \Pi^\alpha_\alpha.             \label{Trace}
\eeq
    Assuming that the right-hand side is a known function of the radial
    coordinate, i.e., that $R = R(r)$ is known, \eq (\ref{Trace}) may be
    written as a linear first-order equation with respect to $f(r) :=
    r B(r)$ \cite{bwh1,vis-wilt}:
\bearr                                                       \label{master}
    A(rA_r + 4A) f_r + [r(2AA_{rr} -A_r^2) + 3 AA_r] f
\nnn \inch
        					= 2A^2 [2 - r^2 R(r)].
\ear
    Its general solution is
\bearr                                                       \label{sol}
	f(r) = \frac{2 A \e^{3\Gamma}}{(4A + rA_r)^2}
\nnn   \cm\
       \times	\int
	    (4A + rA_r) [2 - r^2 R(r)] \e^{- 3\Gamma}\, dr
\ear
    where
\beq                                                          \label{Gamma}
	\Gamma(r) = \int \frac{A_r dr}{4A + rA_r}.
\eeq
    Thus, choosing any smooth function $A(r)$, we obtain $f(r)$ from
    (\ref{sol}), and, after fixing the integration constant, the metric is
    known completely.

    If the function $R(r)$ is not specified, \eq (\ref{sol}) is simply
    another form of the trace of the Einstein equations. It is valid for
    {\it any } \ssph\ metric, at least in ranges of $r$ where the quantity
    $A(4A + rA_r)$ is finite and nonzero and where
    $r = \sqrt{- g_{\theta\theta}}$ is an admissible coordinate. The latter
    means, in particular, that \eq (\ref{sol}) is applicable to a \wh\
    metric only on one (but either) side of a \wh\ throat (see \cite{bwh1}
    for details) and is evidenly invalid for flux-tube metrics,
    characterized by $g_{\theta\theta} =\const$.

    For a given SET $T\mN$, the dependence $R(r)$ is not always known, but
    in the vacuum case $T\mN =0$, so that $R(r) = 4 \Lambda_4$, the solution
    to the Einstein equations can always be written in the form (\ref{sol})
    under the above evident restrictions.

\section {Black hole construction  \label{s3}}

\subsection{Conditions at horizons}    \label{3.1}

    Before singling out BH solutions on the basis of \eq (\ref{sol}), let us
    first formulate the conditions under which the generating function
    $A(r)$ leads to a metric with a Killing horizon. The latter is a
    surface where a timelike or spacelike Killing vector becomes null. To
    describe Killing horizons (to be called horizons for short) in \sph\
    space-times, it is helpful to use the so-called quasiglobal coordinate
    $u$ specified by the condition $g_{tt} g_{uu} = -1$. The
    metric (\ref{ds-r}) is then rewritten in the form
\beq                                                         \label{ds-u}
	ds^2 = \cA(u) dt^2 - \frac{du^2}{\cA(u)}
	                                      - r^2(u) d\Omega^2,
\eeq
    where the variables are connected with those in
    \eq (\ref{ds-r}) as follows:
\bear  \nq
	\cA(u) \eql A(r), \quad r(u) = r,
	\quad \cA(u)\left(\frac{dr}{du}\right)^2 = B(r).     \label{u->r}
\ear

    The reason for using this coordinate is that, in a close neighbourhood
    of a horizon (a sphere where $g_{tt}=0$), it varies like manifestly
    well-behaved Kruskal-like coordinates used for an analytic continuation
    of the metric \cite{cold,vac1}. Using this coordinate, one can ``cross
    the horizons'' preserving the formally static expression for the metric.
    Both $A$ and $r$ must be smooth functions of $u$ near the horizon $u =
    h$, so that
\beq
	\cA(u) \sim (u - h)^k, \qquad                      \label{A,r_hor}
	                r(u) \approx  r_h + \const\cdot(u - h)^s,
\eeq
    where $k = 1,2,\ldots$ is the order of the horizon (the horizon is
    simple if $k=1$, double if $k=2$, etc.) while the number
    $s=1,2,\ldots$ characterizes the possible behaviour of $r(u)$;
    $r_h > 0$ is the horizon radius. (We leave aside possible horizons of
    infinite radius which can in principle appear as well \cite{cold}.)
    Generically but not necessarily one has $s=1$. When $s=1$ (i.e., $dr/du$
    is finite at the horizon), the coordinate $r$ can be used for
    continuing the metric through the horizon on equal grounds with $u$.
    The continuation will be analytic if both $\cA(u)$ and $r(u)$ are
    analytic at $u=h$.

    Assuming (\ref{A,r_hor}) and directly employing the relations
    (\ref{u->r}), we find that
\beq                                                             \label{f1}
    f(r) \sim B(r) \sim (r-r_h)^{(k+2s-2)/s} \quad{\rm as} \quad u\to h.
\eeq

    On the other hand, substituting $A = \cA(u) \sim
    (r-r_h)^{k/s}$ into the solution (\ref{sol}) and assuming that the
    quantity $Q(r) = 2 - r^2 R(r)$ is finite at $r=r_h$ (which is
    generically the case), it is easy to obtain that near $r_h$
\beq                                                        \label{f2}
     f(r) \sim B(r)\sim (r-r_h)^{2-k/s}\left[(r-r_h)^{k/s} + C\right].
\eeq
    where $C$ is an integration constant and $C=0$ corresponds to the case
    when integration in (\ref{sol}) is performed from $r_h$ to $r$.
    Comparing the exponents in (\ref{f1}) and (\ref{f2}), we see that
\begin{itemize}\itemsep 0pt
\item
	in case $C=0$: $k=2$, $s$ is not restricted;
\item
	in case $C\ne 0$: $k=1$, $s$ is not restricted.
\end{itemize}

    Thus, to obtain a solution with a horizon at $r=r_h$, we should take
    $A(r)$ behaving as $(r-r_h)^{k/s}$ with $k=1$ or 2 and $s\in \N$.

    An important point is that, under the condition $R(r_h) \ne 2/r_h^2$,
    a horizon can be either simple ($k=1$) or double ($k=2$); horizons of
    higher orders $k$ do not appear. This is true, in particular, for all
    \ssph\ metrics with $R=0$.

    Let us now look what changes when the function $2 - r^2 R(r)$ vanishes
    at $r=r_h$. One can write
\bearr
    Q(r):= 2- r^2 R(r) \sim (r-r_h)^p,
\nnn
    \cm p=0,1,2,\ldots,  \label{0_p}
	                \quad {\rm near} \quad r=r_h,
\ear
    preserving the assumptions (\ref{A,r_hor}). So $p=0$ corresponds to
    the above generic case $Q(r_h) \ne 0$ and $p>0$ means that $Q(r)$ has a
    zero of order $p$. Then the expression (\ref{f1}) for $f(r)$ remains the
    same but (\ref{f2}) must be replaced with
\beq                                                          \label{f2a}
	f(r) \sim B(r)
	        \sim (r-r_h)^{2-k/s}\left[(r-r_h)^{k/s + p} + C\right]
\eeq
    Consequently, in case $C\ne 0$ the metric behaves as before,
    whereas for $C=0$ we obtain near $r=r_h$
\bearr                                                          \label{high}
	A(r) \sim (r-r_h)^{p+2/s}, \cm
	B(r) \sim (r-r_h)^{p+2},
\nnn
	\cA(u) \sim (u-h)^{ps+2},
\ear
    that is, a horizon of order $ps+2$.

    Now, assuming $A = 0$ at $r=r_h > 0$, one can rewrite
    \eq (\ref{sol}) in the form
\bearr                                                          \label{sol_}
    f(r) \equiv r B(r) = \frac{2 A \e^{3\Gamma}}{(4A + rA_r)^2}
\nnn  \
     \times \Biggl\{\int_{r_h}^r
	    (4A + rA_r) [2 - r^2 R(r)] \e^{- 3\Gamma}\, dr + C\Biggr\}.
\ear

    The above analysis shows that this relation leads to a metric with a
    horizon at $r=r_h$ in two cases:

\begin{description} \itemsep 0pt
\item[(i)]
    $A \sim (r-r_h)^{1/s}$, as $r\to r_h$, $s\in \N$. Then \eq
    (\ref{sol_}) leads to a metric with a simple horizon in case $C \ne 0$
    and a metric with a horizon of order $2 + ps$ in case $C=0$.

\item[(ii)]
    $A \sim (r-r_h)^{2/s}$ as $r\to r_h$, $s$ odd. Then (\ref{sol_})
    leads to a metric with a horizon of order $2 + ps$ in case $C=0$.

\end{description}
    Here, as before, the parameter $p$ characterizes the behaviour of $Q(r)$
    according to \eq (\ref{0_p}).

    Item (ii) does not include solutions with $C\ne 0$. The point is
    that in case $A \sim (r-r_h)^2$, $C \ne 0$, the metric is singular at
    $r = r_h$, as is confirmed by calculating the Kretschmann scalar
    (\ref{Kr}): its constituent $K_2$ blows up at $r \to r_h$. For odd
    numbers $s > 1$, the corresponding metric with $A \sim (r-r_h)^{2/s}$,
    $C \ne 0$ has a finite Kretschmann scalar but loses analyticity at
    $r=r_h$ and therefore cannot be analytically continued through this
    sphere.  Indeed, in this case $r-r_h \sim (u-h)^{s/(ps+2)}$, where the
    exponent is a fraction for any odd $s$ and $p = 0,1,2,\ldots$.
    The metric is thus only continuous ($C^{0}$-smooth) at $u=h$ in case
    $p>0$ and $C^{(s-1)/2}$-smooth in case $p=0$.

\subsection {Definition and algorithms} \label{3.2}

    We have been so far discussing local conditions at possible Killing
    horizons. Let us now turn to space-time properties at large and try to
    select BH metrics. We shall not need a general rigorous definition of a
    BH \cite{wald} which in turn needs such notions as strong asymptotic
    predictability, trapped regions etc. The following working
    definition will be appropriate for our purposes.

\Theorem{Definition}
       {The metric (\ref{ds-u}) is said to describe a \bh\ if (a) the
	functions $\cA(u)$ and $r(u)$ are analytic in the range $\cR[u]: \
	h\leq u < u_{\max}$ where $u_{\max}$ may be finite or infinite; (b)
	$r(u) > 0$ in $\cR[u]$, and $r(u_{\max}) > r(h) = r_h$; (c) $\cA(u)
	> 0$ at $u > h$, and $\cA(h) \sim (u-h)^k,\ k\in \N$ as $r\to r_h$.}

    Item (c) means that $u > h$ is a static region (R region) of a \ssph\
    space-time while the sphere $u = h$, a boundary of this region, is a
    Killing horizon of a certain order $k$. So, in usual terms, our
    working definition describes the domain of outer communication of a BH,
    and $u=h$ is its event horizon.

    The definition uses the $u$ coordinate rather than $r$, due to its
    advantage in horizon description, discussed in the previous subsection.
    The difference is really essential: there are metrics which behave
    non-analytically in terms of $r$ at $r = r(h) = r_h$ but analytically in
    terms of $u$ at $u = h$ (see variants $s > 1$ in \sect \ref{s3} and
    Example 5 in the next section).

    The analyticity requirement rejects possible cases of restricted
    smoothness (see the end of \sect \ref{3.1}). It is not only a matter of
    simplicity: in our view, if we are dealing with a field configuration,
    its non-analyticity at a certain surface must have a physical reason,
    e.g., a phase transition, and it seems too artificial, a kind of perfect
    fine tuning, to assume that the phase transition occurs precisely at
    a horizon.

    We do not require $r(u_{\max}) = \infty$ since we do not want to rule
    out metrics with cosmological horizons like the \Sch-\dS\ space-time
    where an R region is situated between a BH horizon and a cosmological
    horizon. We, however, adopt the requirement $r(u_{\max}) > r_h$ to
    exclude configurations with only cosmological horizons.

    Let us return to the solution (\ref{sol}), or (\ref{sol_}). Due to its
    generality, it certainly describes all BH metrics, at least piecewise.
    We can, however, formulate explicit requirements to the generating
    function $A(r)$ under which \eq (\ref{sol}) leads {\it
    algorithmically\/} to a BH metric.  Namely, let there be a range
\beq
    \cR[r]: \  r_{\max} > r > r_h, \quad r_h > 0,         \label{range}
\eeq
    in which the r.h.s. of \eq (\ref{master}) is positive:
\beq
    	Q(r) = 2- r^2 R(r) > 0.              \label{Q>0}
\eeq
    Then the above items (i) and (ii) lead to the following BH construction
    algorithms.

\Theorem {(BH1)}
       {Specify a function $A(r)$, positive and analytical in $\cR[r]$,
        in such a way that $g(r) = 4A + rA_r > 0$ in $\cR[r]$ and $A \sim
	(r-r_h)^{1/s}$, $s\in \N$, as $r\to r_h$. Then the functions $A(r)$
	and $B(r)$ given by \eq (\ref{sol_}) with $C\geq 0$ determine a \bh\
	metric (\ref{ds-r}) with a horizon at $r=r_h$. The horizon is simple
	if $C > 0$; in case $C=0$ it is of the order $2+ps$ if $Q(r)$ behaves
	according to \eq (\ref{0_p}). }

\Theorem {(BH2)}
       {Specify a function $A(r)$, positive and analytical in $\cR[r]$,
        in such a way that $g(r) = 4A + rA_r > 0$ in $\cR[r]$ and $A \sim
        (r-r_h)^{2/s}$ as $r\to r_h$, $s$ being an odd positive integer. Then
	the functions $A(r)$ and $B(r)$ given by \eq (\ref{sol_}) with $C=0$
	determine a \bh\ metric (\ref{ds-r}) with a horizon at $r=r_h$ of the
	order $2+ps$ if $Q(r)$ behaves according to \eq (\ref{0_p}). }

    Both algorithms (BH1) and (BH2) lead to double horizons in case $C=0$ if
    $Q(r_h) >0$.

    To obtain {\bf \asflat} BHs, one should assume $r_{\max} = \infty$ and
    restrict the choice of $A(r)$ to functions compatible with asymptotic
    flatness. Properly choosing the time scale, we can require a \Sch\
    behaviour of $A$: $A = 1 - 2m/r + o(1/r)$ as $r\to \infty$, where $m$
    is the \Sch\ mass. \eq (\ref{sol_}) then leads to the proper behaviour
    of $B(r)$, i.e., $B \to 1$ as $r\to \infty$, provided the curvature $R$
    decays quickly enough: $R(r) = o(r^{-3})$ as $r\to\infty$.

    One can notice that both algorithms use the $r$ coordinate whereas the
    BH definition uses $u$. A transition to $u$ is accomplished with \eq
    (\ref{u->r}). The conditions of (BH1) and (BH2) guarantee that $B > 0$
    at $r > r_h$, therefore, choosing $dr/du > 0$ in (\ref{u->r}), we
    evidently satisfy the BH definition.

    The condition (\ref{Q>0}) can be weakened: what actually must be
    required is that the integral in (\ref{sol_}) should remain positive in
    $\cR[r]$.

    Another condition of both algorithms, $g(r) = 4A + rA_r > 0$ in $\cR[r]$,
    allows one to avoid a singular point of \eq (\ref{master}), where the
    coefficient of the derivative $f_r$ vanishes. This coefficient also
    vanishes at horizons, which have been already discussed. The points
    where $A \neq 0$ but $g=0$, if any, also deserve special attention.
    Though, the equality $g=0$ itself has no evident geometric (and hence
    physical) meaning, it only represents some technical difficulty in our
    description with the aid of \eq (\ref{master}).

    Points where $g=0$ are avoided by most of \asflat\ BH metrics. Indeed,
    in this case $g \geq 0$ at the horizon and $g =4$ at infinity; the
    condition $g(r) > 0$ means that $r^4 A$ is a strictly increasing
    function of $r$, which is the case, e.g., for all functions $A(r)$
    monotonically growing from 0 at the horizon to 1 at infinity.

    However, one can easily verify that such points inevitably appear in
    other important situations, e.g., between a regular centre and a
    cosmological horizon or between two simple horizons.

    It can be shown that if the choice of the generating function $A(r)$
    leads to $g(r)$ with a simple zero at some $r = r_s \in \cR[r]$, then
    there is a unique nonzero value of the constant $C$ (namely, when
    integration in \eq (\ref{sol}) starts from $r_s$) making it possible to
    avoid a singularity of the metric at $r=r_s$. So Algorithm (BH1)
    survives, but it now gives a single solution with a horizon instead of a
    family parametrized by $C$ (see Example 4 in \sect \ref{s4}). Algorithm
    (BH2) requires $C=0$ and therefore does not work.

\subsection
    {Generic behaviour of the solutions: \whs\ and regular BHs  \label{3.3}}

    Let us discuss the properties of the metric in the generic situation
    that leads to a BH according to Algorithm (BH1): let $A(r)$ be a
    well-behaved positive function at $r > r_h$ and have a simple zero at
    $r=r_h$, and let also $Q (r_h) > 0$. Then, in a small neighbourhood of
    $r=r_h$, one can write
\bear
      A(r) \eql A_1 (r-r_h) + o (r{-}r_h),
\nn                                                         \label{B_hor}
      B(r) \eql B_1 C\,(r-r_h) +B_2 (r{-}r_h)^2 + o\left((r{-}r_h)^2\right),
\ear
    with fixed positive constants $A_1$, $B_1$ and $B_2$. The integration
    constant $C$ is a family parameter, and $C=0$ is its critical value at
    which the solution drastically changes its properties.

    If $C < 0$, $B$ turns to zero at $r = r_{\min} = r_h + |C|B_1/B_2 >r_h$.
    (Here, $|C|$ should be small enough for the solution to remain in a
    range where \eq (\ref{B_hor}) is still approximately valid.) One obtains
    $B(r) \approx B_2(r-r_h)(r-r_{\min})$, so that $B(r)$ has a simple zero
    at $r=r_{\min}$, whereas $A(r_{\min}) > 0$. Such a behaviour of the
    metric functions corresponds to a symmetric \wh\ throat at $r=r_{\min}$
    \cite{bwh1}. The substitution $r = r_{\min} + x^2$, $x \in \R$, makes
    the metric (\ref{ds-r}) regular at $r=r_{\min}$ ($x=0$), and all metric
    coefficients are even functions of $x$. Thus our solution does not reach
    the anticipated horizon $r=r_h$ and describes a symmetric \wh.

    In case $C=0$, as already described, we obtain a double horizon at
    $r=r_h$, and the geometry is smoothly continued to smaller $r$, where
    the further properties of the metric depend on the specific choice of
    $A(r)$.

    If $C > 0$, then $B > 0$ at $r > r_h$, turns to zero at the horizon
    $r=r_h$ and again turns to zero in the T region at $r = r_{\min} = r_h-
    C\, B_1/B_2 < r_h$. This, as before, bounds the range of $r$ from below
    in a way similar to a \wh\ throat. The coordinate singularity at $r=r_0$
    is again removed by the transformation $r = r_{\min} + x^2$, but now
    $x$ is a temporal coordinate in a T region, where the metric (\ref{ds-r})
    describes a Kantowski-Sachs cosmology with two scale factors $r(x)$ and
    $A(r(x))$ and the $\R\times \S^2$ topology of spatial sections. Hence,
    $x=0$ is the time instant at which $r(x)$ experiences a bounce. It can
    be roughly said that the \wh\ throat, having moved into a T region,
    becomes a bouncing time instant of the scale factor $r$ in a
    Kantowski-Sachs cosmology.

\begin{figure}
\centering
\unitlength=0.65mm
\special{em:linewidth 0.5pt}
\linethickness{0.5pt}
\begin{picture}(136.00,100.00)(23,30)
\put(95.00,106.00){\line(1,1){10.00}}
\put(125.00,116.00){\line(1,-1){10.00}}
\put(135.00,106.00){\line(-1,-1){10.00}}
\put(105.00,96.00){\line(-1,1){10.00}}
\put(104.00,96.00){\line(-1,1){10.00}}
\put(94.00,106.00){\line(1,1){10.00}}
\put(126.00,116.00){\line(1,-1){10.00}}
\put(136.00,106.00){\line(-1,-1){10.00}}
\put(105.00,96.00){\line(1,1){20.00}}
\put(105.00,116.00){\line(1,-1){20.00}}
\put(95.00,66.00){\line(1,1){10.00}}
\put(125.00,76.00){\line(1,-1){10.00}}
\put(135.00,66.00){\line(-1,-1){10.00}}
\put(105.00,56.00){\line(-1,1){10.00}}
\put(104.00,56.00){\line(-1,1){10.00}}
\put(94.00,66.00){\line(1,1){10.00}}
\put(126.00,76.00){\line(1,-1){10.00}}
\put(136.00,66.00){\line(-1,-1){10.00}}
\put(105.00,56.00){\line(1,1){20.00}}
\put(105.00,76.00){\line(1,-1){20.00}}
\put(95.00,86.00){\line(1,1){10.00}}
\put(125.00,96.00){\line(1,-1){10.00}}
\put(135.00,86.00){\line(-1,-1){10.00}}
\put(105.00,76.00){\line(-1,1){10.00}}
\put(104.00,76.00){\line(-1,1){10.00}}
\put(94.00,86.00){\line(1,1){10.00}}
\put(126.00,96.00){\line(1,-1){10.00}}
\put(136.00,86.00){\line(-1,-1){10.00}}
\put(105.00,76.00){\line(1,1){20.00}}
\put(105.00,96.00){\line(1,-1){20.00}}
\put(75.00,116.00){\line(1,-1){10.00}}
\put(85.00,106.00){\line(-1,-1){10.00}}
\put(76.00,116.00){\line(1,-1){10.00}}
\put(86.00,106.00){\line(-1,-1){10.00}}
\put(75.00,76.00){\line(1,-1){10.00}}
\put(85.00,66.00){\line(-1,-1){10.00}}
\put(76.00,76.00){\line(1,-1){10.00}}
\put(86.00,66.00){\line(-1,-1){10.00}}
\put(75.00,96.00){\line(1,-1){10.00}}
\put(85.00,86.00){\line(-1,-1){10.00}}
\put(76.00,96.00){\line(1,-1){10.00}}
\put(86.00,86.00){\line(-1,-1){10.00}}
\put(75.00,116.00){\line(-1,-1){10.00}}
\put(65.00,106.00){\line(1,-1){10.00}}
\put(75.00,96.00){\line(-1,-1){10.00}}
\put(65.00,86.00){\line(1,-1){10.00}}
\put(75.00,76.00){\line(-1,-1){10.00}}
\put(65.00,66.00){\line(1,-1){10.00}}
\put(64.00,54.00){\rule{1.00\unitlength}{66.00\unitlength}}
\put(29.00,78.00){\line(1,1){10.00}}
\put(39.00,68.00){\line(-1,1){10.00}}
\put(38.00,68.00){\line(-1,1){10.00}}
\put(28.00,78.00){\line(1,1){10.00}}
\put(39.00,88.00){\line(1,-1){10.00}}
\put(49.00,78.00){\line(-1,-1){10.00}}
\put(40.00,88.00){\line(1,-1){10.00}}
\put(50.00,78.00){\line(-1,-1){10.00}}
\put(38.00,42.00){\makebox(0,0)[cc]{(a): $C<0$}}
\put(75.00,42.00){\makebox(0,0)[cc]{(b): $C=0$}}
\put(115.00,42.00){\makebox(0,0)[cc]{(c): $C>0$}}
\put(39.00,88.00){\line(0,-1){3.00}}
\put(39.00,83.00){\line(0,-1){3.00}}
\put(39.00,78.00){\line(0,-1){3.00}}
\put(39.00,73.00){\line(0,-1){3.00}}
\put(106.00,96.00){\line(1,0){3.00}}
\put(111.00,96.00){\line(1,0){3.00}}
\put(116.00,96.00){\line(1,0){3.00}}
\put(121.00,96.00){\line(1,0){3.00}}
\put(106.00,76.00){\line(1,0){3.00}}
\put(111.00,76.00){\line(1,0){3.00}}
\put(116.00,76.00){\line(1,0){3.00}}
\put(121.00,76.00){\line(1,0){3.00}}
\put(106.00,56.00){\line(1,0){3.00}}
\put(111.00,56.00){\line(1,0){3.00}}
\put(116.00,56.00){\line(1,0){3.00}}
\put(121.00,56.00){\line(1,0){3.00}}
\put(106.00,116.00){\line(1,0){3.00}}
\put(111.00,116.00){\line(1,0){3.00}}
\put(116.00,116.00){\line(1,0){3.00}}
\put(121.00,116.00){\line(1,0){3.00}}
\put(75.00,87.00){\makebox(0,0)[cc]{R}}
\put(36.00,77.00){\makebox(0,0)[cc]{R}}
\put(105.00,87.00){\makebox(0,0)[cc]{R}}
\put(126.00,86.00){\makebox(0,0)[cc]{R}}
\put(115.00,79.00){\makebox(0,0)[cc]{T}}
\put(115.00,92.00){\makebox(0,0)[cc]{T}}
\put(68.00,76.00){\makebox(0,0)[cc]{R}}
\put(68.00,96.00){\makebox(0,0)[cc]{R}}
\end{picture}
\caption
{Carter-Penrose diagrams for a generic family of \asflat\ solutions: (a)
a symmetric \wh, (b) an extremal BH in case there is a singularity inside
the inner R region, and (c) a regular BH. Diagrams (b) and (c) can be
infinitely continued upward and downward. The letters R and T mark static
(R) and cosmological (T) regions, respectively. Spatial infinity
($r=\infty$) is shown by double lines, horizons ($r=r_h$) by single thin
lines and the singularity in (b) by a thick line.  Dashed lines show the
\wh\ throat in diagram (a) and the bouncing time instants in diagram (c).}
\end{figure}
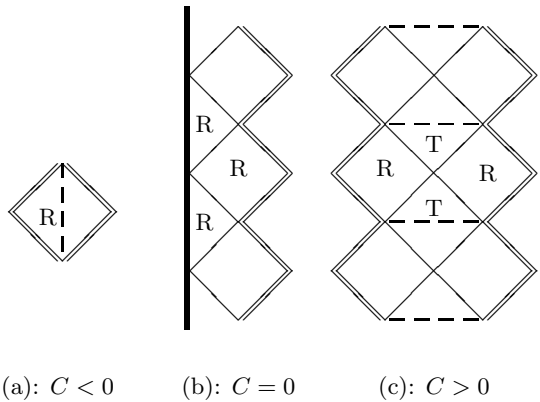

    Assuming asymptotic flatness at large $r$, one can use the standard
    methodology to obtain the corresponding Carter-Penrose diagram
    describing the global causal structure: it coincides with that of the
    Kerr or Kerr-Newman non-extremal BH \cite{wald} (but without a ring
    singularity) and contains an infinite sequence of R and T regions.

    By continuity, this behaviour is preserved in a finite range of $C$
    values (see Examples 1 and 2 in the next section). We conclude that each
    generic choice of $A(r)$ with a simple zero at $r=r_h$ leads to a family
    of solutions unifying symmetric \whs ($C < 0$), extremal BHs ($C=0$) and
    regular non-extremal BHs ($C > 0$), see Fig.\,1.

\section {Examples} \label{s4}

\subsection*{Example 1: $A(r) = 1 - 2m/r$, $m = \const > 0$.}

    For this \Sch\ form of $A(r)$, the solution (\ref{sol}) in the vacuum
    case $R \equiv 0$ can be written as
\beq
	f(r) = \frac{(r-2m)(r-r_0)}{r-3m/2},                  \label{f-x1}
\eeq
    where $r_0$ is an integration constant. The metric takes the form
\bearr
	ds^2 = \schd dt^2                                     \label{ds-x1}
\nnn \qquad
	     - \frac{1 - 3m/(2r)}{(1 - 2m/r)(1 - r_0/r)}dr^2 -r^2 d\Omega^2.
\ear
    The \Sch\ metric is restored in the special case $r_0 = 3m/2$. The
    metric (\ref{ds-x1}) was obtained by Casadio, Fabbri and Mazzacurati
    \cite{casad01} in search for new brane-world \bhs\ and by Germani and
    Maartens \cite{maart01} as a possible external metric of a homogeneous
    star on the brane. Without repeating their more detailed descriptions,
    we will outline the main points in our notations.

    BH metrics appear according to Algorithm (BH1), where $C=0$ corresponds
    to $r_0 = 2m$ and $C > 0$ to $r_0 < 2m$. In case $r_0 > 2m$, the metric
    (\ref{ds-x1}) describes a symmetric traversable \wh\ \cite{bwh1}.

    In case $r_0 = 2m$ we have a double horizon at $r = 2m$: near $r=2m$,
    the coordinate $r$ is connected with the quasiglobal coordinate $u$ by
    $r - 2m \sim (u - u_h)^2$, $u_h$ being the value of $u$ at the horizon,
    and $\cA(u) \sim (u-u_h)^2$. The Carter-Penrose diagram coincides with
    that of the extremal \RN\ metric [Fig.\,1(b)] with the only difference
    that the timelike curvature singularity occurs at $r=3m/2$ instead of
    $r=0$.

    In case $r_0 < 2m$, as in the \Sch\ metric, $r=2m$ is a simple
    horizon, and, as described in Ref.\,\cite{casad01}, the space-time
    structure depends on the sign of $\eta = r_0 - 3m/2$. If $\eta < 0$, the
    structure is that of a \Sch\ \bh, but the spacelike curvature
    singularity is located at $r= 3m/2$ instead of $r=0$. If $\eta > 0$, the
    solution describes a nonsingular \bh\ with a \wh\ throat at $r=r_0$
    inside the horizon, or, more precisely, it is the minimum value of $r$
    at which the model bounces. The corresponding global structure
    \cite{casad01} is the same as that of a non-extremal Kerr BH
    [Fig.\,1(c)].

    Thus the metric properties in the whole range $r_0 > 3m/2$ of the
    integration constant $r_0$ entirely conform to the description in \sect
    \ref{3.3} for both positive and negative $C$.

    The components of the effective SET (\ref{T_eff}) have the form
\bear                                                      \label{T-x1}
    \kappa_4^2 \rho\eff \eql \frac{m(2r_0-3m)} {r^2(2r-3m)^2},
\nn
    \kappa_4^2 \prad\eff \eql -\frac{2r_0-3m} {r^2(2r-3m)},
\nn
    \kappa_4^2 p_\bot\eff \eql \frac{(r-m)(2r_0-3m)} {r^2(2r-3m)^2}.
\ear

\subsection*{Example 2: $A(r) = 1 - h^2/r^2$, $h = \const > 0$.}

    This form of $A(r)$ represents a metric with zero \Sch\ mass.

    BH solutions are easily obtained: \eq (\ref{sol_}) with
    $R\equiv 0$ now gives
\bear
    f(r) = rB(r) = r \biggl(1-\frac{h^2}{r^2}\biggr)          \label{f-x2}
	        \biggl(1 + \frac{C-h}{\sqrt{2r^2 - h^2}}\biggr).
\ear
    In accord with (BH1), the sphere $r=h$ is a simple horizon if $C >0$ and
    a double horizon if $C=0$.

    In case $C < 0$, $B(r)$ has a simple zero at $r =r_{\rm th} > h$ given by
\beq
    2 r^2_{\rm th} = h^2 + (h - C)^2,                       \label{th-x2}
\eeq
    which is a symmetric \wh\ throat \cite{bwh1}.

    In case $C=0$, $r=h$ is a double horizon, and the Carter--Penrose
    diagram coincides with that of the extremal \RN\ metric [Fig.\,1(b)],
    but a timelike singularity due to $B \to \infty$ takes place at $r =
    h/\sqrt{2}$.

    In case $0 < C < h$, inside the simple horizon $r=h$, the function
    $B(r)$ turns to zero at $r=r_{\rm th}$ given by (\ref{th-x2}), which is
    now between $h$ and $h/\sqrt{2}$, and we obtain a Kerr-like regular BH
    structure with an infinite sequence of R and T regions [Fig.\,1(c)]. We
    see that the description of \sect \ref{3.3} is valid in the whole range
    $C < h$ of the integration constant $C$.

    The value $C = h$ leads to the simplest metric with $A = B = 1 -
    h^2/r^2$, which may be identified as the \RN\ metric with zero mass and
    pure imaginary charge. The space-time causal structure is \Sch, with
    a horizon at $r=h$ and a singularity at $r=0$. Lastly, in case $C > h$
    the causal structure is again \Sch\ but the singularity due to $B \to
    \infty$ occurs at $r=h/\sqrt{2}$.

    This example is of certain interest in connection with Thorne's
    ``hoop conjecture'', claiming that a BH horizon forms when and only
    when a mass $M$ gets concentrated in a region whose circumference in
    every direction is smaller than $4\pi GM$, $G$ being Newton's constant
    of gravity \cite{hoop}. Nakamura et al. \cite{hoop-b} recently found an
    example of a cylindrical (i.e., infinitely long) matter distribution on
    the brane able to form a horizon and thus violating the hoop conjecture.
    The present example of a zero mass BH shows that, in the \bwd\ context,
    a BH may exist (at least as a solution to the gravitational equations on
    the brane) without matter and without mass, solely as a tidal effect
    from the bulk gravity. The effective SET is in this case certainly
    quite exotic from the viewpoint of the conventional energy conditions:
\bear                                                      \label{eff-x2}
     \kappa_4^2 \rho\eff \eql
		-\frac{h^2}{r^4} - \frac{h^2(C-h)(3r^2-h^2)}
			{r^4(2r^2-h^2)^{3/2}},
\nn
     \kappa_4^2 \prad\eff \eql
		\frac{h^2}{r^4} + \frac{(C-h)(r^2+h^2)}
			{r^4(2r^2-h^2)^{1/2}},
\nn
     \kappa_4^2 p\bot\eff \eql
	       -\frac{h^2}{r^4} - \frac{(C-h)(r^4 + 2h^2 r^2 - h^4)}
			{r^4(2r^2-h^2)^{3/2}}.
\ear
    In the simplest case $C=h$ it has the ``anti-\RN'' form,
    $\propto r^{-4}\diag (-1,\,-1,\,1,\,1)$.

\subsection*{Example 3: $A(r) = (1 - 2m/r)^2$, $m = \const > 0$.}

    For this extremal \RN\ form of $A(r)$, the solution (\ref{sol}) with $R
    \equiv 0$ and the metric can be written as
\bear
	f(r) \eql \frac{(r-r_0)(r-r_1)}{r},
	        \cm        r_1 := \frac{mr_0}{r_0-m},        \label{f-x3}
\\
    ds^2 \eql \schd^2 dt^2                                   \label{ds-x3}
\nnn \ \
		-\biggl(1-\frac{r_0}{r}\biggr)^{-1}
	        \biggl(1-\frac{r_1}{r}\biggr)^{-1} dr^2 -r^2 d\Omega^2.
\ear

    The form of $A(r)$ fits to Algorithm (BH2), and accordingly we obtain a
    BH solution in the only case $r_0 = r_1 = 2m$, i.e., when the
    integration in \eq (\ref{sol_}) is conducted from $r_h = 2m$, so
    that $C=0$. It is the extremal \RN\ metric, and accordingly the effective
    SET is $T\mN{}\eff \propto r^{-4}\diag (1,\,1,\,-1,\,-1)$.

    Other values of $r_0$ lead either to \whs\ (the throat is located at
    $r = r_0$ if $r_0 > 2m$ or at $r = r_1 > 2m$ in case $2m > r_0 > m$), or
    to a naked singularity located at $r=2m$ (when $r_0 < m$) as is
    confirmed by calculating the Kretschmann scalar --- see more detail in
    Ref.\,\cite{bwh1}.

\subsection*{Example 4: $A(r) = 1 - r^2/a^2$, $a = \const > 0$.}

    The above examples described vacuum \asflat\ BHs.
    Now, choosing the de Sitter form of $A(r)$, we will write the solution
    (\ref{sol}) for a vacuum configuration with a cosmological term, so
    that $R = 4 \Lambda_4 = 12/a^2$, in the region $r < a$. We obtain
\beq
	f(r) = rB(r) = \biggl(1 - \frac{r^2}{a^2}\biggr)         \label{f-x4}
	\biggl[r + \frac{K}{(2a^2 - 3r^2)^{3/2}} \biggr],
\eeq
    where $K$ is an integration constant such that $K=0$ corresponds to
    integration in \eq (\ref{sol}) from $r = r_s = a \sqrt{2/3}$ to $r$.
    The value $r=r_s$ is the one where $g(r) = 4A + rA_r$ vanishes.
    In full agreement with the description in \sect \ref{3.2}, $B(r)$ tends to
    infinity as $r\to r_s$ unless $K=0$, and thus the only well-behaved
    solution is de Sitter, with $A = B = 1 - r^2/a^2$. This example
    illustrates what happens when \eq (\ref{master}) has a singular point
    $g=0$ in the range of interest.

\subsection*
    {Example 5: $A(r) = (1 - 2m/r)^{1/s}$, $m = \const > 0$, $s\in \N$}

    We here try to give an example of a metric behaving non-analytically
    at $r=r_h$ in terms of $r$ but analytically in terms of the quasiglobal
    coordinate $u$ defined by $g_{tt} g_{uu}=-1$, see \sect \ref{3.1}. A
    certain difficulty is that the solution (\ref{sol}) for this choice of
    $A(u)$, even in the simplest case $R=0$, is expressed with the aid of the
    hypergeometric function, which can hardly be a very clear illustration.
    We therefore simply take the following ``artificial'' example of an
    \asflat\ metric (\ref{ds-r}):
\beq                                                           \label{x5a}
	ds^2 = \schd^{1/s} dt^2 - \schd^{-2+1/s}dr^2 -r^2 d\Omega^2,
\eeq
    as suggested by \eq (\ref{f1}) for any positive integer $s$ in the case
    of a simple horizon, and transform it to a coordinate which behaves like
    $u$ near $r_h=2m$, namely, put $1-2m/r = x^s$ (we do not directly use
    the transformation (\ref{u->r}) since $u(r)$ then looks too cumbersome.)
    The metric takes the form
\beq                                                            \label{x5b}
	ds^2 = x dt^2 - \frac{4m^2 s^2}{x(1-x^s)^4}dx^2
	    			- \frac{4m^2}{(1-x^s)^2} d\Omega^2.
\eeq
    Its asymptotic flatness at $x =1$ is not so evident, but evident is the
    behaviour at $x=0$ as expected at a simple horizon. In case $s=1$ it
    is the \Sch\ metric. For $s >1$ it is not a vacuum solution to \eq
    (\ref{master}); the effective SET is easily found according to \eqs
    (\ref{T_eff}), it decays at large $r$ as $r^{-4}$, in particular, its
    trace is $\kappa_4^{-2}$ times the scalar curvature
\[
	R = \frac{2}{r^2} + \frac{2}{r^3 s}\schd^{1-1/s}[2m-s(r+2m)].
\]

    The same substitution $1 - 2m/r = x^s$ applied to the metric
\beq                                                          \label{x5c}
	ds^2 = \schd^{2/s} dt^2 - \schd^{-2}dr^2 -r^2 d\Omega^2
\eeq
    with $m>0$ and $s\in \N$ reveals a double horizon at $r=2m$.

\section {Concluding remarks} \label{s5}

    Using the trace of the 4D Einstein equations, written as a linear
    first-order ordinary differential equation and integrated, we have
    formulated some general requirements to the (arbitrary) generating
    function $A(r) \equiv g_{tt}$ which are sufficient for obtaining \ssph\
    BH metrics --- see Algorithms (BH1) and (BH2). The latter may be
    \asflat\ or have any other large $r$ behaviour.

    We have seen that, under some natural restrictions, BH metrics are
    easily constructed in vacuum or in the presence of matter for which the
    dependence $R(r)$ may be specified. Though, not every kind of matter
    distribution admits a horizon inside it. No horizon can appear, e.g., in
    a perfect fluid with the equation of state $\rho = np,\ n\in \N$: the
    conservation law then implies $\rho = \rho_0 A^{-(n+1)/2}$, $\rho_0 =
    \const$, so that $\rho \to \infty$ as $A\to 0$. More generally, at a
    horizon, the effective SET (\ref{T_eff}) should satisfy the condition
    $\rho\eff + \prad\eff = 0$. Indeed, one can write in terms of the metric
    (\ref{ds-u}):
\beq                                                          \label{01}
    G^t_t - G^u_u
    	= 2\,\frac{\cA}{r} \frac{d^2 r}{du^2}
    		= -\kappa_4^2 (\rho\eff + \prad\eff)
\eeq
    [recall that $\cA(u) = A(r)$], which leads to $\rho\eff + \prad\eff = 0$
    at regular points where $\cA =0$.

    The same quantity is negative at \wh\ throats (it is the well-known
    violation of the null energy condition \cite{hoh-vis}) but is positive at
    bounces in T regions. \eq (\ref{01}) shows that this property is quite
    general: at a minimum of $r$, where $d^2 r/du^2 > 0$, one has
    $\rho\eff + \prad\eff < 0$ if $\cA > 0$ (a throat) and
    $\rho\eff + \prad\eff > 0$ if $\cA < 0$ (a bounce).

    A feature of utmost interest is the generic appearance of families of
    solutions which unify symmetric \whs\ and globally regular BHs with a
    bounce of $r$ in the T region and a Kerr-like global structure. The
    two qualitatively different branches of any such family are separated by
    an extremal BH solution.

    Certain care should be taken about possible zeros ($r=r_s$) of the
    function $g(r) = 4A + rA_r$ which is a coefficient of the derivative
    $f_r$ in \eq (\ref{master}). We have shown that even if the choice of
    $A(r)$ leads to $g(r) = 0$ at some $r$, a well-behaved solution can
    generically be obtained.

    The whole consideration is quite general and may find application in
    \GR\ (where our effective SET ${T\mN}\eff$ is simply the matter SET
    including a possible cosmological term) and alternative theories of
    gravity that use modified Einstein equations. As usual, for a particular
    kind of matter, the theory will give one more independent equation for
    the two metric functions to be found, $A(r)$ and $B(r)$ in our notation,
    and the set of equations will be determined.

    The most natural application of these results is, however, the \bw\
    concept where the trace of the Einstein equations is the only equation
    of 4D gravity which can be written unambiguously using only 4D
    quantities.  Being a single equation for the two unknown functions
    $A(r)$ and $B(r)$, it leads to a variety of BH as well as \wh\
    solutions. This ambiguity reflects the ambiguity of brane embedding into
    the bulk, which manifests itself in \eqs (\ref{EE4}) in the
    arbitrariness of $E\mN$. Moreover, as remarked in Ref.\,\cite{bwh1},
    the tensor $E\mN$, due to its geometric origin, need not respect the
    usual energy conditions, and the appearance of \wh\ solutions in its
    presence looks quite natural.

    The above arbitrariness exists even in the simplest \bw\ models of RS2
    type \cite{ransum2}, possessing a single extra dimension, $\Z_2$
    symmetry with respect to the brane and no matter in the bulk, to say
    nothing of more complex models. The latter may include scalar fields in
    the bulk \cite{od-bh,sc-bulk}, multiple (at least two) branes
    \cite{ransum1,2bra}, more than one extra dimension \cite{guen}, timelike
    extra dimensions, lacking $\Z_2$ symmetry, a 4D curvature term
    \cite{shta} etc; see further references in the cited papers and the
    reviews \cite{bra-rev}. To improve the predictive power of \bw\
    scenarios, it seems necessary to remove the ``redundant freedom'',
    applying reasonable physical requirements such as regularity and
    stability to complete multidimensional models.

    Much work in this direction has already been done. Different methods of
    solving the bulk gravity equations for given brane configurations have
    been developed \cite{chamb01,wiseman,casad02,vis-wilt}, and the bulk
    properties of some particular \bwd\ BHs have been studied [2--5, 8]
    as well as their possible quantum properties \cite{casad-qu}. It appears
    to be a necessary, though difficult, task to extend the study to more
    general BH configurations.

\subsection*{Acknowledgments}

{The authors acknowledge partial financial support from DFG project
        436/RUS 113/678/3-1 and also (KB and VM) from RFBR Grant
	01-0217312a. KB and VM are thankful to the colleagues from the
	University of Konstanz for kind hospitality.}

\end{document}